# Reply to Comment by V. P. Torchigin and A. V. Torchigin on "Deducing radiation pressure on a submerged mirror from the Doppler shift"

Masud Mansuripur

College of Optical Sciences, The University of Arizona, Tucson, Arizona 85721
masud@optics.arizona.edu

The criticism of V. P. Torchigin and A. V. Torchigin appears to be directed at my application in [1] of a result obtained from classical (Maxwellian) electrodynamics to the case of a single photon of frequency $\omega_0$ and energy $\hbar\omega_0$. Having derived a solution for the classical problem of reflection of a light pulse from a submerged mirror in Sec.V of [1], I suggested that the same result is applicable to individual photons provided that the total energy of the light pulse is reduced until it reaches the energy $\hbar\omega_0$ of a single photon. This is not unlike the standard treatment of Young's double-slit experiment, in which a single photon passes through both slits, with the resulting probability amplitudes interfering with each other at the observation plane. The classical interference fringe pattern in Young's double-slit experiment is subsequently recovered when the experiment is repeated with a large number of single photons successively passing through the same pair of slits.

Another relevant example is the splitting of the probability amplitude of a single photon at the first beam-splitter of a Mach-Zehnder interferometer, followed by a recombination of the amplitudes at the second beam-splitter. Once again, the classical result is recovered when a large number of photons successively pass through the interferometer and their emergence probabilities from each output channel are compared with the intensities of classical light beams emerging from the same two channels.

The connection between the above examples and the analysis associated with Fig.1 of my paper [1] is that the probability amplitude of a single photon reaching the rear facet of the dielectric slab at $z = 0$ splits between a reflected path and a transmitted path, followed by a superposition of these amplitudes when the transmitted wave returns from the mirror located at $z = d$, having acquired a phase $\psi_0 = \psi + 4\pi d/\lambda_0$, which is imparted by the mirror and by propagation within the gap. In other words, the single photon cannot be assumed to have returned *either* from the rear facet of the dielectric slab or from the mirror, without accounting for the interference between the corresponding probability amplitudes. The consequences of this interference, of course, are implicit in the classical electromagnetic analysis presented in Sec.V of [1]. It is noteworthy that this classical result is also in full agreement with the single-photon calculations based on the Doppler shift of the reflected photon, as reported in Sec. IV of [1].

The Torchigins proceed to calculate the intensity of a classical light beam in the gap between the dielectric slab and the mirror in order to point out that the $\psi$-dependence of the radiation pressure felt by the mirror is a consequence of the tuning or detuning of the resonator formed between the mirror and the rear facet of the slab. These authors apparently missed the exact same calculation that I presented in Sec.V of [1], where my gap-field $E_g$ is the same as their transmitted field amplitude $a_2$, my Eq.(10) is the same as their Eq.(8), and their assertions following their Eq.(8) are implicit in my Eq.(11), where $\psi_0 = \psi + 4\pi d/\lambda_0$ is given by Eq.(3c) of [1]. According to my Eq.(11), the pressure on the mirror is proportional to $n$ when $\psi_0 = \pi$, and to $1/n$ when $\psi_0 = 0$, in agreement with the Torchigins' contention. More importantly, however, the Torchigins seem to ignore the main point of these calculations, namely, that the aforementioned proportionality to $n$ (or to $1/n$) persists when the gap is closed (i.e., $d \to 0$), in which case the photons that bombard the mirror no longer cross a free-space gap, but rather arrive at the mirror's surface directly through the dielectric medium of refractive index $n$.

In their second paragraph, the Torchigins state: "*In reality, a number of photons reflected from the mirror per unit time depends on the phase angle $\psi_0$ and may vary in the interval from $N/n$ to $Nn$ where $N$ is the number of photons per unit time in the incident light pulse. The momentum of a single photon reflected from the mirror does not depend on the phase angle $\psi_0$ and is equal to the momentum of a photon in free space $p_0 = \hbar\omega_0/c$.*" I do not quite know what to make of this statement. All I can say is that, with the gap between the mirror and the dielectric slab closed, the number of photons per unit time arriving at the mirror will be equal to $N$, which is the number of photons per unit time passing through any cross-section of the incident light pulse. Moreover, as stated in [1], the force of radiation felt by the mirror will be somewhere in the interval between $2N\hbar\omega_0/(nc)$ and $2nN\hbar\omega_0/c$, depending on the phase angle $\psi$ of the mirror. Nowhere in [1] have I suggested that the momentum of a single photon reflected from the mirror in any way depends on the phase angle $\psi$ of the mirror. In fact, I have stated explicitly (in Sec. I and again in Sec. VI) that the photon momentum inside the dielectric slab of refractive index $n$ is equal to $\frac{1}{2}(n + n^{-1})\hbar\omega_0/c$.

Finally, a confusing aspect of the Comment by the Torchigins is that the authors use a non-standard form for the Fresnel reflection and transmission coefficients. In their Eqs.(1)-(5), the symbols $a$ and $b$ are meant to represent the amplitude of the EM field. Apparently, the authors do not equate the field amplitude with either the electric field $E$ or the magnetic field $H$ of the light wave. The standard expressions for the reflected and transmitted $E$-field amplitudes are given by the Fresnel reflection and transmission coefficients, namely, $\rho = E^{(r)}/E^{(i)}$ and $\tau = E^{(t)}/E^{(i)}$ [where the superscripts $(i)$, $(r)$, and $(t)$ stand for incident, reflected, and transmitted], as follows:

$$\rho_{\text{air-to-glass}} = (1 - n)/(1 + n), \tag{1a}$$

$$\tau_{\text{air-to-glass}} = 2/(1 + n), \tag{1b}$$

$$\rho_{\text{glass-to-air}} = (n - 1)/(n + 1), \tag{1c}$$

$$\tau_{\text{glass-to-air}} = 2n/(n + 1). \tag{1d}$$

Clearly, $\rho^2 + \tau^2 \neq 1$ for both cases of air-to-glass and glass-to-air incidence, in contrast to the Torchigins' assertion in their Eq.(7) that $t^2 + r^2 = 1$. Conservation of energy, however, requires careful attention to the magnitude of the $H$-field. Denoting the impedance of free space by $Z_0 = \sqrt{\mu_0/\varepsilon_0}$, where $\mu_0$ and $\varepsilon_0$ are, respectively, the permeability and permittivity of free space, the magnetic-field amplitude of a plane-wave in an isotropic, homogeneous, linear dielectric can be shown to be $H_0 = nE_0/Z_0$. Subsequently, the rate of flow of EM energy along the $z$-axis is given by the time-averaged Poynting vector, $\langle \mathbf{S} \rangle = \langle \mathbf{E} \times \mathbf{H} \rangle = \frac{1}{2}(nE_0^2/Z_0)\hat{\mathbf{z}}$, which, in conjunction with the Fresnel reflection and transmission coefficients given by the above Eqs.(1), satisfies the requirement of energy conservation.

1. M. Mansuripur, "Deducing radiation pressure on a submerged mirror from the Doppler shift," Phys. Rev. A **85**, 023807 (2012).